\documentclass{IEEEtran4PSCC}
\ifCLASSINFOpdf
   \usepackage[pdftex]{graphicx}
\else
   \usepackage[dvips]{graphicx}
\fi
\usepackage[cmex10]{amsmath}
\makeatletter
\let\old@ps@headings\ps@headings
\let\old@ps@IEEEtitlepagestyle\ps@IEEEtitlepagestyle
\def\psccfooter#1{%
    \def\ps@headings{%
        \old@ps@headings%
        \def\@oddfoot{\strut\hfill#1\hfill\strut}%
        \def\@evenfoot{\strut\hfill#1\hfill\strut}%
    }%
    \def\ps@IEEEtitlepagestyle{%
        \old@ps@IEEEtitlepagestyle%
        \def\@oddfoot{\strut\hfill#1\hfill\strut}%
        \def\@evenfoot{\strut\hfill#1\hfill\strut}%
    }%
    \ps@headings%
}
\makeatother

\psccfooter{%
        \parbox{\textwidth}{\hrulefill \\ \small{24th Power Systems Computation Conference} \hfill \begin{minipage}{0.2\textwidth}\centering \vspace*{4pt} \includegraphics[scale=0.06]{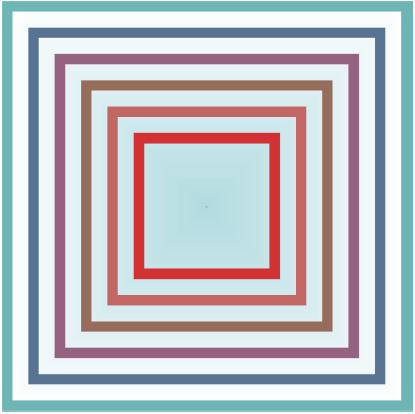}\\\small{PSCC 2026} \end{minipage} \hfill \small{Limassol, Cyprus --- June 8-12, 2026}}%
}
\usepackage{amsmath,amsfonts}
\usepackage{amsthm}
\usepackage{balance}

\newtheoremstyle{ieeetheorem}
{}                % Space above
{}                % Space below
{\upshape}        % Theorem body font % (default is "\upshape")
{}                % Indent amount
{\bfseries}       % Theorem head font % (default is \mdseries)
{.}               % Punctuation after theorem head % default: no punctuation
{ }               % Space after theorem head
{}                % Theorem head spec
\theoremstyle{ieeetheorem}

\theoremstyle{definition}

\usepackage{bm}
\usepackage{algorithm}
\usepackage[noend]{algpseudocode}
\usepackage{graphicx}
\usepackage{color}
\usepackage{array}
\usepackage{url}
\usepackage{cite}
\usepackage{booktabs}
\usepackage{multirow}
\usepackage{multicol}
\usepackage{hyperref}
\usepackage{tikz}
\usepackage{tikz-3dplot}            % 3D coordinates support
\usepackage{pgfplots}                % plotting
\usepgfplotslibrary{fillbetween}     % fill between plots
\pgfplotsset{compat=1.18}            % use latest compatible version

\usetikzlibrary{
    arrows.meta,                      % improved arrows
    patterns, patterns.meta,          % patterns support
    decorations, decorations.text, decorations.pathreplacing, 
    shapes, positioning, shadows, backgrounds, calc
}

\usepackage{fontawesome}              % icons
\usepackage{xcolor}                   % colors
\definecolor{maincolor}{HTML}{00274C}
\definecolor{green}{RGB}{34,139,34}
\definecolor{red}{RGB}{156,31,46}
\definecolor{deep_blue}{RGB} {0,0,255}
\makeatletter
\newcommand{\oset}[3][0ex]{%
  \mathrel{\mathop{#3}\limits^{
    \vbox to#1{\kern-2\ex@
    \hbox{$\scriptstyle#2$}\vss}}}}
\makeatother

\newcommand{\mb}[1]{\mathbf{#1}}
\newcommand{\mbg}[1]{\boldsymbol{#1}}

\begin{document}
%
% paper title
% Titles are generally capitalized except for words such as a, an, and, as,
% at, but, by, for, in, nor, of, on, or, the, to and up, which are usually
% not capitalized unless they are the first or last word of the title.
% Linebreaks \\ can be used within to get better formatting as desired.
% Do not put math or special symbols in the title.

\title{DiffOPF: Diffusion Solver for Optimal Power Flow}

%% To specify the authors when (number of affiliations <= 2)
\author{
\IEEEauthorblockN{Milad Hoseinpour, Vladimir Dvorkin}
\IEEEauthorblockA{Department of Electrical Engineering and Computer Science\\
University of Michigan,\\
Ann Arbor, USA\\
\{miladh, dvorkin\}@umich.edu}
}

\maketitle

\begin{abstract}
The optimal power flow (OPF) is a multi-valued, non-convex mapping from loads to dispatch setpoints. The variability of system parameters (e.g., admittances, topology) further contributes to the multiplicity of dispatch setpoints for a given load. Existing deep learning OPF solvers are single-valued and thus fail to capture the variability of system parameters unless fully represented in the feature space, which is prohibitive.  To solve this problem, we introduce a diffusion-based OPF solver, termed \textit{DiffOPF}, that treats OPF as a conditional sampling problem. The solver learns the joint distribution of loads and dispatch setpoints from operational history, and returns the marginal dispatch distributions conditioned on loads. Unlike single-valued solvers, DiffOPF enables sampling statistically credible warm starts with favorable cost and constraint satisfaction trade-offs. We explore the sample complexity of DiffOPF to ensure the OPF solution within a prescribed distance from the optimization-based solution, and verify this experimentally on power system benchmarks.
\end{abstract}


% Use this to place sponsorships
%\thanksto{\noindent Submitted to the 24th Power Systems Computation Conference (PSCC 2026).}

\begingroup
\allowdisplaybreaks

\section{Introduction}
The optimal power flow (OPF) problem, as a nonconvex and nonlinear optimization, is a computationally challenging task. The integration of renewables on the supply side, and data centers on the demand side, further increase variability and uncertainty in power systems, forcing system operators to solve OPF problems more frequently in response to constantly changing parameters \cite{khaloie2025review,liang2024generative, swirydowicz2024gpu, tang2017real}. To address time-sensitive conditions, deep learning models have been proposed as computationally efficient alternatives to traditional optimization-based solvers, producing solutions in a fraction of the time \cite{van2021machine}.

The seminal works in \cite{zamzam2020learning,fioretto2020predicting,pan2020deepopf} have laid the foundations of deep learning solutions to OPF. Given the load input, the solver in \cite{zamzam2020learning} yields an efficient warm start for a non-convex OPF problem, significantly reduces the computational burden. Since then, substantial research efforts have focused on improving the performance of load-to-dispatch mappings, with emphasis on operational and security constraints satisfaction \cite{fioretto2020predicting,pan2022deepopf,velloso2021combining,min2024hard}, integrating extra domain knowledge \cite{singh2021learning}, resolving uncertainties \cite{gupta2022dnn}, compressing the mapping \cite{10246400,kody2022modeling}, and developing topology-informed architectures for deep learning OPF solvers \cite{liu2022topology,wu2025universal}. The interested reader is referred to \cite{ML_OPF_wiki} for a comprehensive list of contributions in this area. 

\begin{figure*}[t]
    \centering
    \includegraphics[width=0.85\linewidth]{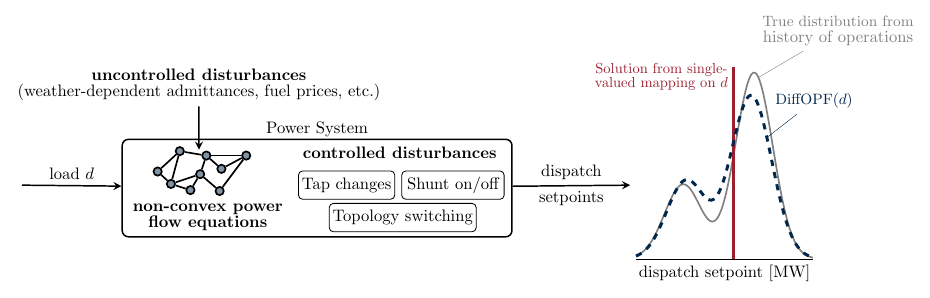}

    \vspace{-0.25cm}
    \caption{\textit{Multi-valued load-to-dispatch mapping in OPF.} The multiplicity of dispatch setpoints in OPF arises from several sources: the inherent non-convexity of power flow equations, uncontrolled disturbances affecting OPF parameters, and controlled disturbances, such as transformer tap changes, shunt switching, and network topology switching. As a result, for a given load input $d$, the history of operations yields a distribution of dispatch setpoints. Trained on historical observations, single-valued deep learning OPF solvers return a deterministic prediction for $d$, typically the conditional expectation. DiffOPF, instead, learns the full distribution of dispatch setpoints, returning a credible population of setpoints consistent with the historical observations.}
\label{fig:intro}
\end{figure*}

However, the inherent complexities of the load-to-dispatch OPF mapping pose practical challenges for single-valued deep learning OPF solvers, as depicted in Fig. \ref{fig:intro}. First, the underlying OPF mapping admits multiple global and local solutions \cite{molzahn2019survey}, which cannot be fully captured by single-valued solvers unless provided by extra information. This problem has been addressed by work in \cite{pan2023deepopf}, where the authors augmented the feature space with initialization for power injection and voltage variables. Second, power system operation is influenced by controlled and uncontrolled disturbances, both contributing to the stochastic nature of each operational record. The former disturbances relate to constantly changing transformer tap settings, shunt and topology switching, and corrective actions. The latter disturbances arise from weather- and market-induced variability of OPF parameters, such as temperature-sensitive admittances, cost functions affected by fuel prices, and market strategies among others. While some sources of variability, such as network parameters or topology, can in principle be incorporated in the design of deep learning OPF solvers (e.g., by expanding the feature space~\cite{pan2023deepopf} or integrating grid topology~\cite{liu2022topology,wu2025universal}), accounting for all operational factors in this way is generally infeasible. Moreover, associating each operational record with the precise realization of these factors is often not possible due to incomplete or unavailable data. 

In this paper, we develop a data-driven solver that addresses the multi-valued nature of the load-to-dispatch mapping in OPF. The mapping is abstracted by a diffusion model, termed \textit{DiffOPF}, whose output is the conditional distribution of generator setpoints given load inputs. Sampling from the conditional distribution allows system operators to generate warm-start points for downstream power flow problems, enabling efficient exploration of dispatch options and cost vs. constraint satisfaction trade-offs. Our contributions include:
\begin{enumerate}
    \item We cast the OPF problem as a sampling problem, following the principles of diffusion and inverse problems in machine learning \cite{ho2020denoising,song2020score,chung2022diffusion}. The diffusion model learns the full load–dispatch distribution from the history of operations. During sampling, given load input, the model serves as a prior to generate statistically credible dispatch setpoints supported by operational history. 

    \item The conditional sampling enables system operators to quantify uncertainty in dispatch setpoints and generate multiple statistically credible warm starts for a given load level. Each sample provides distinct cost and constraint satisfaction outcomes, allowing operators to select solutions that best align with current operation and risk tolerance. Thus, DiffOPF improves on single-valued deep learning OPF solvers in warm start applications.
    
    \item We establish the sample complexity of DiffOPF, providing a lower bound on the number of samples required to generate a solution within a prescribed distance from the solution of conventional interior-point solvers.  
\end{enumerate}

More broadly, our work aligns with the emerging concept of decision flows, which aims to sample decisions from target distributions guided by prior samplers \cite{chertkov2025sampling}. Preliminary applications of decision flows with custom diffusion processes to unit commitment appear in \cite{ferrando2025physics}. In this work, however, we leverage established DDPM \cite{ho2020denoising} and score-based diffusion models \cite{song2020score} rather than developing custom diffusion processes. 

While other generative models such as GANs and VAEs have been explored in power systems, their applications have primarily focused on generating synthetic data, such as load or renewable generation profiles \cite{fekri2019generating,pan2019data}, rather than addressing inverse problems arising in system operation. Moreover, in GAN-based approaches, constraints are typically incorporated through penalization in the training objective, which can exacerbate training instability and mode collapse \cite{yang2020physics,jabbar2021survey}. In contrast, diffusion models provide a principled mechanism for incorporating constraints during sampling. This capability forms the foundation of DiffOPF, enabling the generation of dispatch solutions conditioned on demand levels, which act as structural constraints in the inverse problem. 

Prior work \cite{hoseinpour2025constraineddiffusionmodelssynthesizing} demonstrated that DDPMs are well suited for OPF-related applications, enabling control of generated samples to satisfy power flow constraints in the synthesis of power flow datasets. Although extending such constraint enforcement to data-driven OPF solvers is a natural next step \cite{fioretto2020predicting,pan2022deepopf,min2024hard}, this paper focuses on establishing the foundational principles of DiffOPF for warm-start applications, leaving full constraint satisfaction for future work.

\section{Problem Setup}

We frame the OPF problem in a data-driven setting. The inputs to the OPF problem are the loads and grid data, e.g., network topology, line parameters, and the outputs are the optimal dispatch setpoints. Let $\mb{p}_d \in \mathbb{R}^{n_d}, \; \mb{q}_d \in \mathbb{R}^{n_d}$ denote the vector of active and reactive power demands, and $\mb{p}_g \in \mathbb{R}^{n_g}, \; \mb{q}_g \in \mathbb{R}^{n_g}$ denote the vector of active and reactive dispatch setpoints. We assume access to a dataset $\mathcal{D} = \{(\mb{p}_{d}^{i}, \mb{q}_{d}^{i}, \mb{p}_{g}^{i}, \mb{q}_{g}^{i}) \}_{i=1}^{N}$, where each record corresponds to an OPF solution under given operating conditions. Concatenating the variables into a joint vector $\mb{x}_0=(\mb{p}_{d},\mb{q}_{d},\mb{p}_{g},\mb{q}_{g})$, we view the OPF dataset $\mathcal{D}$ as samples from a true distribution $p_{\text{real}}(\mb{x}_0)$ of operational records. DiffOPF has two stages:
\begin{enumerate}
    \item Learn a generative model $p_{\theta}(\cdot)$ to approximate $p_{\text{real}}(\cdot)$;
    \item Once the model is learned, produce a conditional distribution $p_{\theta}(\mb{p}_g,\mb{q}_g|\mb{p}_d,\mb{q}_d)$ for a given load input $(\mb{p}_d, \mb{q}_d)$.
\end{enumerate}

Sampling from the conditional distribution produces statistically credible warm start solutions for the downstream power flow problem. An overview of this setup is given in Fig.~\ref{fig:schematic}.  

In the next Sec. \ref{sec:prelimianries}, we introduce modeling preliminaries on AC-OPF and diffusion models in the AC-OPF context. Then, in Sec. \ref{sec:diffusion_guidance}, we introduce our approach for conditional sampling.

\begin{figure*}[t]
    \centering
    \includegraphics[width=0.85\linewidth]{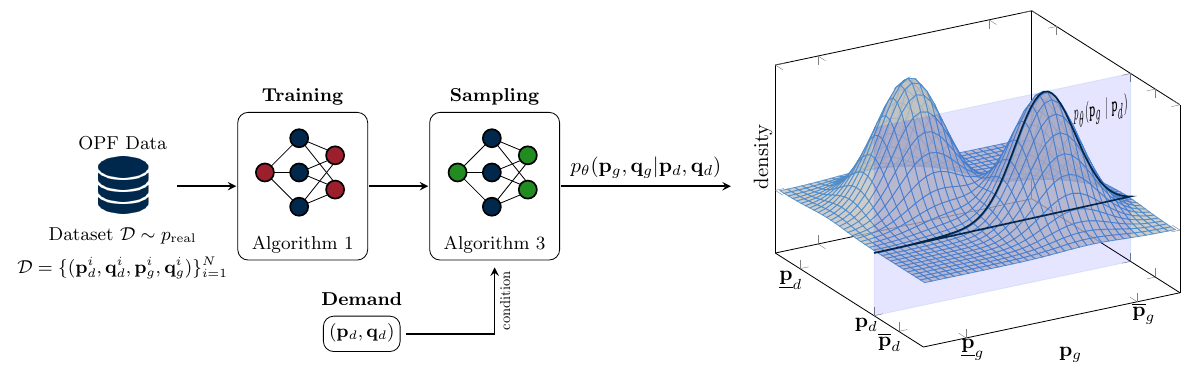}

    \vspace{-0.25cm}
\caption{Schematic of the proposed DiffOPF framework. Historical OPF records in $\mathcal{D}$ are used to learn the joint distribution $p_{\text{real}}(\mathbf{p}_d, \mathbf{q}_d,\mathbf{p}_g, \mathbf{q}_g)$. At the next stage, given a new demand observation $(\mathbf{p}_d, \mathbf{q}_d)$, the model generates OPF solutions $(\mathbf{p}_g, \mathbf{q}_g)$ by sampling the conditional distribution $p_{\theta}(\mathbf{p}_g, \mathbf{q}_g|\mathbf{p}_d, \mathbf{q}_d)$. In the interest of illustration, we only show a two-dimensional joint distribution between $\mathbf{p}_d$ and $\mathbf{p}_g$.}
\label{fig:schematic}
\end{figure*}

\section{Modeling Preliminaries}\label{sec:prelimianries}
\subsection{AC Optimal Power Flow Problem}

The OPF problem seeks to determine generator dispatch setpoints that minimize operational costs while satisfying the power flow and grid constraints \cite{chatzivasileiadis2018lecture}. Let $\mathcal{B}={1,\ldots,B}$ denote the set of buses and $\mathcal{L}={1,\ldots,L}$ the set of transmission lines. For each bus $b \in \mathcal{B}$, let $p_{g,b}$ and $q_{g,b}$ denote the active and reactive power dispatch, and $p_{d,b}$ and $q_{d,b}$ denote the active and reactive power demand, respectively. The bus voltage magnitude and angle are represented by $v_b$ and $\theta_b$. For brevity, shunt admittances are omitted from the formulation, although they are accounted for in the numerical experiments. The non-linear, non-convex AC-OPF problem takes the form:
\begin{subequations}
\label{detACOPF1}
\begin{align}
&\underset{\mb{p}_g,\mb{q}_g,\mb{v},\mbg{\theta}}{\text{minimize}}\quad c(\mb{p}_{g})\label{obj}\\
&\text{subject to}\nonumber\\
% generation constraints
&p_{g,b}-p_{d,b}-\sum_{l \in \mathcal{L}: i=b} f^p_{l, i\to j}-\sum_{l \in \mathcal{L}: j=b} f^p_{l, j\to i}=0, \label{ac_powerbal}  \\
&q_{g,b}-q_{d,b}- \sum_{l \in \mathcal{L}: i=b} f^q_{l, i\to j} - \sum_{l \in \mathcal{L}: j=b} f^q_{l, j\to i}=0, \label{reac_powerbal}\\
&f^p_{l,i\to j} =  v_i v_j \big[g_{l} \cos (\theta_i - \theta_j) + b_{l}  \sin (\theta_i - \theta_j)\big], \label{ac_pf_eq1p}\\
&f^q_{l,i\to j}= v_i v_j \big[g_{l} \sin(\theta_i - \theta_j) - b_{l} \cos (\theta_i - \theta_j)\big], \label{ac_pf_eq1q}\\
&\underline{p}_{g,b}\leq p_{g,b} \leq \overline{p}_{g,b}, \label{ac_p}\\
&\underline{q}_{g,b} \leq q_{g,b} \leq \overline{q}_{g,b}, \label{ac_q}\\
% voltage constraint
&\underline{v}_{b} \leq v_{b} \leq \overline{v}_{b}, \label{ac_v}\\
%&\theta_{slack} = 0 &&\label{ac_ref} 
&(f^p_{l,i\to j})^2+ (f^q_{l,i\to j})^2 \leq(\overline{s}_{l})^2.\label{ac_s} 
\end{align}
\end{subequations}

The objective function \eqref{obj} minimizes the total generation costs. For each bus $b\in\mathcal{B}$, Eq. \eqref{ac_powerbal}–\eqref{reac_powerbal} are the active and reactive power balance constraints, and Eq. \eqref{ac_p}–\eqref{ac_q} and \eqref{ac_v} enforce the generator and voltage limits, respectively. For each transmission line $l \in \mathcal{L}$ from node $i$ to $j$, the active $f^{p}_{l,i\to j}$ and reactive $f^{q}_{l,i\to j}$ power flows are given by \eqref{ac_pf_eq1p}–\eqref{ac_pf_eq1q}, and the apparent power flow limit is enforced by \eqref{ac_s}. 

\subsection{Diffusion Models}
Diffusion models are generative models that synthesize new data through a two-stage process: forward and reverse. Consider $\mb{x}_0 =(\mb{p}_{d}, \mb{q}_{d}, \mb{p}_{g}, \mb{q}_{g})$ as a real OPF data point drawn from the underlying distribution $p_{\text{real}}$ of OPF records. The forward process gradually perturbs a clean data point $\mathbf{x}_0 \sim p_{\text{real}}$ into a noisy sample $\mathbf{x}_T \sim p_T$, where $p_T$ is a standard multivariate Gaussian distribution $\mathcal{N}(0, \mathbf{I})$ with an identity covariance matrix. This can be modeled as the solution to an It\^o stochastic differential equation (SDE) \cite{chung2022diffusion}:
\begin{equation}
    d\mathbf{x}_t = -\frac{\beta(t)}{2}\,\mathbf{x}_t\,dt + \sqrt{\beta(t)}\,d\mathbf{w}_t, 
    \quad t \in [0, T],
    \label{eq:forward_sde_general}
\end{equation}
where $\beta(t) > 0$ is a prescribed noise schedule. For practical implementation, the continuous-time SDE in \eqref{eq:forward_sde_general} is discretized into $T$ steps with a discrete noise schedule $\{\beta_i\}_{i=1}^T$ and corresponding $\alpha_t = 1 - \beta_t$, $\bar{\alpha}_t = \prod_{i=1}^t \alpha_i$. A clean sample $\mathbf{x}_0$ can then be transferred to a noisy sample $\mathbf{x}_t$ at step $t$ via
\begin{equation}
    \mathbf{x}_t = \sqrt{\bar{\alpha}_t}\,\mathbf{x}_0 + \sqrt{1 - \bar{\alpha}_t}\,\mbg{\epsilon}, \qquad \mbg{\epsilon} \sim \mathcal{N}(0, \mathbf{I}).
    \label{discretized_forward}
\end{equation}

The reverse process recovers the underlying data distribution $p_{\text{real}}$ from the noise distribution $p_T$ by the reverse-time SDE:
\begin{equation}
    d\mathbf{x}_t = \big[ -\frac{\beta(t)}{2}\mb{x}_t - \beta(t) \nabla_{\mathbf{x}_t} \log p_t(\mathbf{x}_t) \big] dt 
    + \sqrt{\beta(t)}\, d\bar{\mathbf{w}}_t,
    \label{eq:reverse_sde}
\end{equation}
where $\bar{\mathbf{w}}_t$ is a standard Wiener process running backward in time, and $\nabla_{\mathbf{x}_t} \log p_t(\mathbf{x}_t)$ is the score function of $p_t(\mathbf{x}_t)$. 

The score $\nabla_{\mathbf{x}_t} \log p_t(\mathbf{x}_t)$ in \eqref{eq:reverse_sde} is unknown but can be learned by a neural network $\mbg{s}_\theta(\mathbf{x}, t)$ using denoising score matching \cite{song2020score}. For stability and easier implementation, it is standard to exploit the relationship between noise and score by training a noise predictor $\mbg{\epsilon}_\theta(\mathbf{x}, t)$ in place of the score $\mbg{s}_\theta(\mathbf{x}, t)$ \cite{song2020score}:
\begin{equation}
 \mbg{s}_\theta(\mathbf{x}, t) = -\,\frac{\mbg{\epsilon}_\theta(\mathbf{x}, t)}{\sqrt{1-\bar\alpha_t}}.
\end{equation}
\begin{algorithm}[b]
\caption{:~Training the diffusion model}
\label{alg1}
\textbf{Inputs}: initialized neural network $\mbg{\epsilon}_{\theta}$, noise schedule $\{\alpha_{t}\}_{t=1}^{T}$, dataset of OPF records $\mathbf{x}_0$ from $p_{\text{real}}$\\
\textbf{Outputs}: trained neural network $\mbg{\epsilon}_{\theta}$
\begin{algorithmic}[1]
\Repeat

    \State $\mathbf{x}_0 \sim p_{\text{real}}(\mathbf{x}_0)$
    \State $t \sim \text{Uniform}(\{1, \dots, T\})$
    \State ${\mbg{\epsilon}} \sim \mathcal{N}(0, \mathbf{I})$
    \State Take gradient descent step on
    \[
    \nabla_\theta \left\| {\mbg{\epsilon}} - \mbg{\epsilon}_\theta\left( \sqrt{\bar{\alpha}_t} \mathbf{x}_0 + \sqrt{1 - \bar{\alpha}_t} {\mbg{\epsilon}}, t \right) \right\|^2
    \]
\Until{converged}
\end{algorithmic}
\end{algorithm}
Accordingly, the neural network $\mbg{\epsilon}_\theta(\mathbf{x}, t)$ is trained with a diffusion loss $\mathcal{L}_{\text{diff}}$ defined as the mean-squared error between the actual noise $\mbg{\epsilon} \sim \mathcal{N}(0, \mathbf{I})$ added during the forward process and the predicted noise $\mbg{\epsilon}_\theta(\cdot,t)$:
\begin{equation}
\mathcal{L}_{\text{diff}} = \mathbb{E}_{\mb{x}_0, \mbg{\epsilon}, t} 
\left\|\mbg{\epsilon} - \mbg{\epsilon}_\theta\!\left(\sqrt{\bar{\alpha}_t}\,\mb{x}_0 + \sqrt{1 - \bar{\alpha}_t}\,\mbg{\epsilon},\, t\right) \right\|^2.
\label{loss_function}
\end{equation}
The training procedure is summarized in Alg.~\ref{alg1} \cite{ho2020denoising}. Once the neural network is trained, new OPF data points can be generated by first estimating the clean sample $\hat{\mathbf{x}}_0$ at each step $t$ using the score predictor $\mbg{s}_\theta(\mathbf{x}_t, t)$:
\begin{equation}
\hat{\mathbf{x}}_0= \frac{1}{\sqrt{\bar{\alpha}_t}} \left( \mathbf{x}_t +({1 - \bar{\alpha}_t}) \mbg{s}_{\theta}(\mathbf{x}_t, t) \right),
\label{sampling_1}
\end{equation}
and then by discretizing the reverse-time SDE \eqref{eq:reverse_sde} as
\begin{equation} 
\mathbf{x}_{t-1} = \frac{\sqrt{\alpha_t}(1 - \bar{\alpha}_{t-1})}{1 - \bar{\alpha}_t} \mathbf{x}_t + \frac{\sqrt{\bar{\alpha}_{t-1}}\beta_t}{1 - \bar{\alpha}_t} \hat{\mathbf{x}}_0 + \sigma_t \,\mathbf{z},
\label{sampling_2}
\end{equation}
where $\mb{z}\sim\mathcal{N}(0, \mathbf{I})$, $\sigma_t = \beta_t (1 - \bar{\alpha}_{t-1}) / (1 - \bar{\alpha}_t)$, and $t$ decreases from $T$ (i.e., a pure Gaussian noise) to $1$ (i.e., generated sample). Algorithm~\ref{alg2} summarizes the implementation of the standard sampling process \cite{ho2020denoising}. 
\begin{algorithm}[t]
\caption{:~Sampling OPF records}
\label{alg2}
\textbf{Inputs}: trained score $\mbg{s}_{\theta}$, noise schedule $\{\alpha_{t}\}_{t=1}^{T}$ and scale $\sigma_t$\\
\textbf{Outputs}: synthesized OPF record $\mathbf{x}_0$
\begin{algorithmic}[1]
\State $\mathbf{x}_T \sim \mathcal{N}(0, \mathbf{I})$
\For{$t = T, \dots, 1$}
    \State $\hat{\mathbf{x}}_0 \gets \frac{1}{\sqrt{\bar{\alpha}_t}} \left( \mathbf{x}_t +({1 - \bar{\alpha}_t}) \mbg{s}_{\theta}(\mathbf{x}_t, t) \right)$
    \State $\mathbf{z} \sim \mathcal{N}(0, \mathbf{I})$ if $t > 1$, else $\mathbf{z} = 0$
    \State $\mathbf{x}_{t-1}\gets\frac{\sqrt{\alpha_t}(1 - \bar{\alpha}_{t-1})}{1 - \bar{\alpha}_t} \mathbf{x}_t + \frac{\sqrt{\bar{\alpha}_{t-1}} \beta_t}{1 - \bar{\alpha}_t} \hat{\mathbf{x}}_0 + {\sigma}_t \mb{z}$
\EndFor
\State \textbf{return} $\mathbf{x}_0$
\end{algorithmic}
\end{algorithm}

\section{Solving OPF with Guided Diffusion}\label{sec:diffusion_guidance}

\subsection{DiffOPF Solver}

The diffusion model introduced in previous section learns the distribution of historical OPF records, and then samples from the entire distribution. The model is thus yet to be adapted as a solver to select the correct conditional distribution of dispatch setpoints given loads. Toward conditional sampling, we relate AC-OPF to inverse problems. In general, an inverse problem seeks to recover an unknown variable $\mb{x}_0$ from observations $\mb{y}$ related through a known operator $\mb{A}$, typically expressed as
\begin{equation}
\label{inverse_problem}
\mb{y} = \mb{A} \mb{x}_0.
\end{equation}

In the OPF context, $\mb{x}_0=(\mb{p}_{d},\mb{q}_{d},\mb{p}_{g},\mb{q}_{g})$ is the full operational record, while $\mb{y}$ is the subset of $\mb{x}_0$ collecting the active and reactive power demands. The binary selection matrix $\mb{A} \in \{0,1\}^{2n_d\times2(n_d+n_g)}$ extracts $\mb{y}$ from $\mb{x}_0$. Thus, each OPF record provides the demand vector as the known variable and dispatch setpoints as unknown. The goal of the inverse problem is then to recover unkown dispatch setpoints from known demand inputs.

\begin{algorithm}[tb]
\caption{: Sampling OPF warm starts with guidance}
\label{alg3}
\textbf{Inputs}: $\mbg{s}_{\theta}$, noise schedule $\{\alpha_{t}\}_{t=1}^{T}$ and scale $\sigma_t$, guidance scale $\lambda$, load $\mb{y}=(\mb{p}_d, \mb{q}_d)$\\
\textbf{Outputs}: dispatch setpoint $(\mb{p}_g, \mb{q}_g)$ conditioned on $\mb{y}$
\begin{algorithmic}[1]
\State $\mathbf{x}_T \sim \mathcal{N}(0,  \mathbf{I})$
\For{$t = T-1$ to $0$}
    \State $\hat{\mathbf{x}}_{0} \gets \frac{1}{\sqrt{\bar{\alpha}_t}} \left( \mathbf{x}_t +({1 - \bar{\alpha}_t}) \mbg{s}_{\theta}(\mathbf{x}_t, t) \right)$
     \State $\hat{\mathbf{x}}_{0}^{\prime} \gets \hat{\mathbf{x}}_{0} - \color{red}{\lambda A^{T}(\mb{y}-A\hat{\mb{x}}_0)}$ \Comment{guidance term}
    \State $\mb{z} \sim \mathcal{N}(0,  \mathbf{I})$
    \State $\mathbf{x}_{t-1} \gets \frac{\sqrt{\alpha_t}(1 - \bar{\alpha}_{t-1})}{1 - \bar{\alpha}_t} \mathbf{x}_t + \frac{\sqrt{\bar{\alpha}_{t-1}} \beta_t}{1 - \bar{\alpha}_t}\hat{\mathbf{x}}'_{0} + \sigma_t \mb{z}$
\EndFor
\State\textbf{return} $(\mb{p}_g, \mb{q}_g)\subset\mathbf{x}_0$
\end{algorithmic}
\end{algorithm}
We solve this inverse problem using diffusion. We adopt the Bayesian framework, where the diffusion model serves as a prior distribution $p_{\theta}(\mb{x}_0)$ over OPF records $\mb{x}_0$. During sampling, given an observation $\mb{y}=(\mb{p}_d, \mb{q}_d)$, we sample from the posterior $p_{\theta}(\mb{x}_0|\mb{y})$ to complete the missing entries of record $\mb{x}_0$. For posterior sampling, we adapt \eqref{eq:reverse_sde} as follows:
\begin{equation} 
d\mb{x}_t=\left[-\beta(t)\left(\frac{\mb{x}_t}{2}+\nabla_{\mb{x}_{t}} \log p_t(\mb{x}_{t}|\mb{y})\right)\right]dt+\sqrt{\beta(t)}d\bar{\mathbf{w}}_t,
\label{eq:conditional_posterior}
\end{equation}
where we now have a conditional score $\nabla_{\mb{x}_{t}} \log p_t(\mb{x}_{t}|\mb{y})$, decomposed into two terms using the Bayes’ rule:  
\begin{equation}
\label{conditional_score}
\nabla_{\mb{x}_{t}} \log p_t(\mb{x}_{t}|\mb{y}) =\nabla_{\mb{x}_{t}} \log p_t(\mb{x}_{t}) \;+\; \nabla_{\mb{x}_{t}} \log p_t(\mb{y}|\mb{x}_{t}),
\end{equation}
where the first term corresponds to the pre-trained score function $\boldsymbol{s}_{\boldsymbol{\theta}}$ of the diffusion prior $p_{\theta}(\mb{x}_0)$, and the second term is the guidance term that steers the sampling trajectory toward samples consistent with observed demands~\cite{young2024diffusionposteriorsamplingsequential}. Since there is only an explicit dependence of $\mb{x}_0$ on $\mb{y}$ in \eqref{inverse_problem}, computing the guidance term $\nabla{\mb{x}_{t}}\log p_t(\mb{y}|\mb{x}_t)$ in a closed form is generally intractable. Following ~\cite{chung2022diffusion}, this term is approximated as
\begin{equation}
\nabla_{\mb{x}_{t}}\log p_t(\mb{y}|\mb{x}_t) \approx\nabla_{\mb{x}_{t}}\log p_t(\mb{y}|\hat{\mb{x}}_0),
\end{equation}
where $\hat{\mb{x}}_0$ is the estimate of the solution ${\mb{x}}_0$ at step $t$.
Now, the guidance term  reduces to
\begin{equation}
\nabla_{\mb{x}_{t}} \log p_t(\mb{y}|\hat{\mb{x}}_0) = \nabla_{\mb{x}_{t}} \log\delta(\mb{y} - A \hat{\mb{x}}_0),
\end{equation}
where $\delta(\cdot)$ is the Dirac delta function. It is common and convenient to replace $\delta(\cdot)$ with a Gaussian likelihoods that concentrate on Eq. \eqref{inverse_problem}. Concretely, for $\sigma_g>0$, we define
\begin{equation}\label{gauss_approx}
p_{t,\sigma_g}(\mb{y}|\hat{\mb{x}}_0)=\mathcal{N}(\mb{y}\mid A\hat{\mb{x}}_0,{\sigma_g}^2 \mathbf{I}_{2n_d})
\propto\exp\Big(-\frac{\| \mb{y}-A\hat{\mb{x}}_0\|_2^2}{2\sigma_g^2}\Big),
\end{equation}
so that $\delta(\mb{y}-A\hat{\mb{x}}_0)=\lim_{{\sigma_g}\rightarrow 0} p_{t,\sigma_g}(\mb{y}|\hat{\mb{x}}_0)$. By using \eqref{gauss_approx}, we obtain a well-defined guidance term:
\begin{align}
\nabla_{\mb{x}_t}\log p_{t,\sigma_g}(\mb{y}|\hat{\mb{x}}_0)
&= \nabla_{\mb{x}_t}\Big[-\frac{1}{2\sigma_g^2}\|\mb{y}-A\hat{\mb{x}}_0\|_2^2)\Big]\nonumber\\
&= \lambda A^{T}(\mb{y}-A\hat{\mb{x}}_0),\label{gauss_guidance}
\end{align}
where $\lambda = 1/\sigma_g^2$ is the scale parameter for guidance, which can be empirically tuned for performance. 

Algorithm~\ref{alg3} incorporates this guidance term into the reverse diffusion process, steering the sampling trajectory toward dispatch solutions consistent with the observed demands. However, the sampling procedure remains stochastic. In particular, two sources of randomness are present: the initial noise $\mb{x}_T$ used to start the reverse process at $t=T$, and the noise $\mb{z}$ injected at each reverse step. These stochastic components lead to different sampling trajectories, allowing the model to explore multiple high-probability regions of the learned conditional distribution. Hence, when historical data contain multiple dispatch patterns associated with similar load levels, different noise realizations can produce diverse candidate dispatch solutions conditioned on the same load.

In the present work, the guidance term only enforces consistency with the specified demand levels and does not explicitly incorporate the OPF constraints during sampling. Therefore, the generated samples are not guaranteed to be fully feasible OPF solutions. Instead, they should be interpreted as candidate dispatch points that can serve as warm-starts for conventional OPF solvers.

Beyond warm-start applications, learning the conditional distribution of generator dispatch given load levels enables several additional use cases. For example, it can help characterize the generation patterns associated with a given load realization, generate diverse candidate operating points for contingency and security assessment, and support uncertainty-aware operational studies. These capabilities arise from modeling a distribution of dispatch solutions rather than producing a single deterministic prediction.

\subsection{Sample Complexity of DiffOPF}\label{subsec:sample_comp}
Conventional deep learning OPF solvers learn a deterministic mapping $f: (\mathbf{p}_d, \mathbf{q}_d) \mapsto (\mathbf{p}_g, \mathbf{q}_g)$ trained to minimize the mean squared error of dispatch setpoints, given load inputs. As a result, they output the conditional mean of dispatch setpoints, rather than modeling their full conditional distribution.

In contrast, the proposed DiffOPF in Alg. \ref{alg3} aims to approximate the full conditional distribution $p(\mb{p}_g, \mb{q}_g|\mb{p}_d, \mb{q}_d)$. Formally, given demand $(\mb{p}_d, \mb{q}_d)$, the solver generates candidate solutions by sampling the conditional distribution $p(\mb{p}_g, \mb{q}_g \mid \mb{p}_d, \mb{q}_d)$. If historical OPF records contain multiple dispatch solutions associated with similar load levels, these appear as distinct modes in the conditional distribution $p(\mb{p}_g,\mb{q}_g|\mb{p}_d,\mb{q}_d)$. Since diffusion models learn the full data distribution, conditional sampling allows the model to generate samples around these modes, thus providing a distributional characterization of dispatch setpoints rather than reducing the solution space to mean values as in single-valued deep learning OPF solvers.

In this section, we determine the number of samples required to obtain an $\varepsilon$-close solution to that provided by conventional interior-point optimization solvers, where $\varepsilon>0$ is a specified performance requirement. Let $\{(\mathbf{p}_g^{i}, \mathbf{q}_g^{i})\}_{i=1}^{M} \sim p(\mathbf{p}_g, \mathbf{q}_g \mid \mathbf{p}_d, \mathbf{q}_d)$ 
be $M$ independent samples from DiffOPF conditioned on a given demand $(\mathbf{p}_d, \mathbf{q}_d)$. Define
\begin{equation}
p_\varepsilon = \Pr\!\left(c(\mathbf{p}_g^{i}) \le c^\star + \varepsilon \,\Big|\, \mathbf{p}_d, \mathbf{q}_d\right)
\end{equation}
as the probability that a single sample is $\varepsilon$-close to the locally OPF solution $c^\star$. By independence, the probability that none of the $M$ samples is $\varepsilon$-close to $c^\star$ is $(1 -p_\varepsilon)^M$. The independence assumption is justified because each sample is generated by an independent run of the diffusion sampler. Due to Alg.~\ref{alg3}, the reverse diffusion process is initialized from an independent Gaussian noise realization and evolves with independent stochastic updates.

Taking the complement gives the probability that at least one sample has this property:
\begin{equation}
\Pr\left(\min_{1 \le i \le M} c(\mb{p}_g^{i}) \le c^\star + \varepsilon\right) = 1 - (1 - p_\varepsilon)^M.
\label{prob._at_least}
\end{equation} 

Due to \eqref{prob._at_least}, to have at least one $\varepsilon$-close solution for a desired confidence level $\delta \in (0,1)$, we require
\begin{equation}
1 - (1 - p_\varepsilon)^M \ge \delta \quad \implies \quad (1 - p_\varepsilon)^M \le 1 - \delta,
\end{equation}
which yields the lower bound on sample complexity
\begin{equation}\label{bound}
M \ge \frac{\log(1 - \delta)}{\log(1 - p_\varepsilon)}.
\end{equation}

With $p_\varepsilon \downarrow$ (smaller $\varepsilon$) and $\delta \uparrow$ (higher confidence), the number of samples increases. In practice, $p_{\varepsilon}$ can be estimated by sampling and computing the fraction of $\varepsilon$-close outcomes, enabling system operators to determine the number of samples for a specified accuracy and confidence. In Sec. \ref{sec:res}, we show that $M = 32$ samples suffice to get a solution within $0.5\%$ of the optimal cost with confidence $90\%$ in the PJM test case. 

% for a given confidence level and $\varepsilon$, for instance $\varepsilon=0.5\%$ and confidence level 0.9, 

\section{Evaluation of DiffOPF Warm-Starts}

We propose DiffOPF for warm-start applications, where the solver provides an initialization for a downstream projection problem. In this regard, for each generated sample $(\mb{p}_g, \mb{q}_g)$ we solve the following problem \cite{fioretto2020predicting}: 
\begin{subequations}
\label{eq:pp_opt}
\begin{align}
\underset{\tilde{\mb{p}}_g, \tilde{\mb{q}}_g, \tilde{\mb{v}}, \tilde{\boldsymbol{\theta}}}{\text{minimize}}\quad&\| \tilde{\mb{p}}_g - \mb{p}_g\|_2^{2} + \| \tilde{\mb{q}}_g - \mb{q}_g\|_2^{2} \label{eq:pp_obj}\\
\text{subject to}\quad&
\text{Power Flow Equations}\;\eqref{ac_powerbal}-\eqref{ac_pf_eq1q},
\end{align}
\end{subequations}
where the initial values for dispatch setpoints are set to the output of DiffOPF, and those for voltage variables are set to nominal quantities. The goal here is to find the closest dispatch setpoints that would satisfy the AC power flow equations. 

In our evaluation, we aim to identify the closest point on the power flow manifold; the better the warm start, the closer this point will be. This also means that enforcing inequality constraints on voltage variables would alter this distance. Therefore, we intend to explore various warm starts without enforcing these constraints and explicitly quantify any resulting constraint violations. Specifically, we use the following metrics for evaluating DiffOPF warm starts:
\begin{enumerate}
\begin{subequations}\label{metrics}
\item Optimality gap 
\begin{equation}\label{metrics:cost}
\left\lVert c(\tilde{\mb{p}}_{g}^{\star}) - c(\mb{p}_{g}^{\star})\right\lVert_2,
\end{equation}
defined as the cost difference between the projection \eqref{eq:pp_opt} solution and the dispatch from the historical dataset.

\item Dispatch setpoint error
\begin{equation}\label{metrics:dispatch}
\left\lVert\tilde{\mb{p}}_{g}^{\star} - \mathbf{p}_g\right\lVert_2^{2} + \left\lVert\tilde{\mb{q}}_{g}^{\star} - \mathbf{q}_g\right\lVert_2^{2},
\end{equation} 
as the distance from warm-start to feasible dispatch.

\item Magnitude of voltage constraint violations, measured as  
\begin{equation}\label{metrics:voltage}
\left\lVert \max(0, \tilde{\mathbf{v}}^{\star} - \overline{\mathbf{v}})\right\lVert_2 + \left\lVert \max(0, \underline{\mathbf{v}} - \tilde{\mathbf{v}}^{\star}) \right\lVert_2.
\end{equation}
\end{subequations}
\end{enumerate}

\section{Numerical Experiments}\label{sec:res}

We conduct experiments with DiffOPF to assess the quality of approximating the multi-valued AC-OPF mapping, benchmark its performance against a single-valued deep learning OPF solver, and empirically verify its sample complexity.

\subsection{Experimental Setup}
We generate OPF datasets using benchmark power systems from \cite{babaeinejadsarookolaee2019power}, by solving many instances of the AC-OPF problem. For each system, we sample loads $(\mb{p}_{d}, \mb{q}_{d})$ from a uniform distribution with the range of $80$-$100\%$ of nominal loads, while preserving their power factors. To model uncontrolled disturbances that further contribute to the multiplicity of the load-to-dispatch mapping (see Fig. \ref{fig:intro}),  we randomize the linear cost coefficients by sampling them from a uniform distribution with the range of $\pm40\%$ of nominal values. We obtained $5,000$ samples of training data for each benchmark system.

The multi-valued mapping of DiffOPF is benchmarked against a single-valued map, referred to as DeepOPF after the pioneering work in \cite{pan2020deepopf}. The two solvers have the same input-output relationship and are trained on the same datasets, but DeepOPF maps loads to dispatch setpoints deterministically.

Working with DiffOPF, for each load instance, we sample 250 dispatch setpoints for the projection problem \eqref{eq:pp_opt}, and then report the results for the best sample in terms of each metric in \eqref{metrics}. The sampling time across all test cases does not exceed $2.3$ seconds per DiffOPF sample. When implemented on distributed computing nodes, all samples can be obtained at once, taking $2.3$ seconds in total. 

\subsection{Illustrating Multi-Valued Mapping and Trade-Offs}

To illustrate the ability of DiffOPF to capture the multi-valued load-to-dispatch OPF mapping, we visualize dispatch setpoints in the PJM $5$-bus power system in Fig. \ref{fig:DiffOPF_vs_DeepOPF}. We pick a single test load instance, for which we displayed the historical dispatch records in gray. Observe that the marginal distributions with respect to active power dispatch of generators $3$ and $4$ are multimodal, with two distinguished modes depicted in gray. We then pass the same load input to the two solvers. DeepOPF, as a single-valued solver, only captured the conditional expectation of the two marginal dispatch distributions (depicted in red), while DiffOPF successfully captured their modality and support (depicted in blue).

The capacity of DiffOPF to describe the modes and support of the underlying historical distribution makes it suitable to explore various trade-offs offered by sampled warm starts. The trade-off between the optimality gap and voltage violations on the same load instance in the PJM system is depicted in Fig. \ref{fig:VoltageVsOptGap}. For this particular test sample, the warm start provided by DeepOPF resulted in both moderate voltage violations and optimality gap (yellow cross). In contrast, DiffOPF provided a population of 250 warm starts: many of such warm starts proved to be very unfavorable for dispatch cost and voltage constraint violations. Yet, the majority of warm starts resulted in OPF outcomes concentrated in the favorable southwest part of the plot, with smaller optimality gaps and constraint violations. System operators can pick samples that satisfy their operational objectives: the one with the minimum dispatch cost (red star) or with the lowest voltage violations (green diamond). Either of the two outperform the dispatch outcome associated with the DeepOPF warm start.

\begin{figure}
    \centering
    \includegraphics[width=1\linewidth]{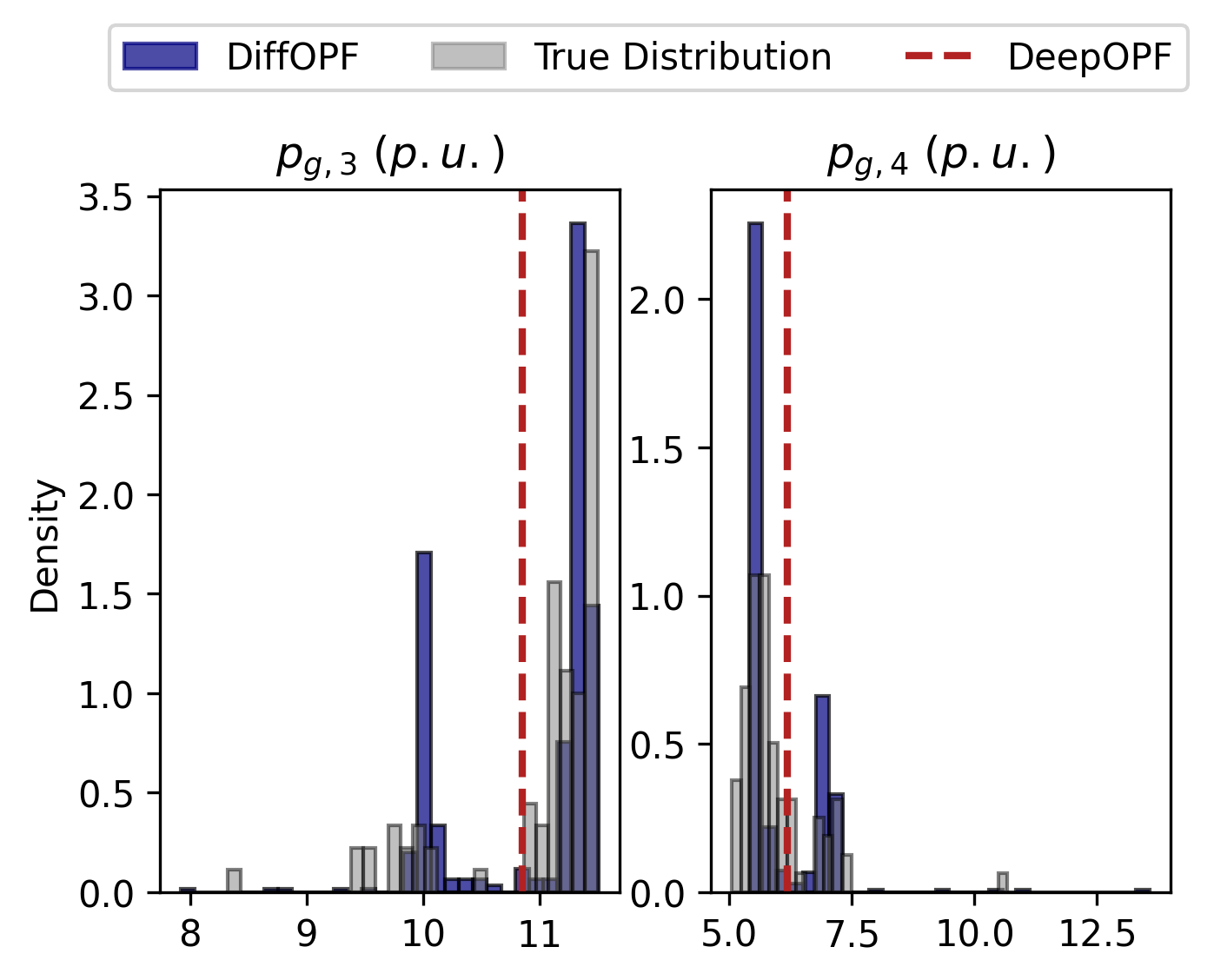}
    \caption{Comparison of DiffOPF and DeepOPF for the selected dispatch setpoints in the PJM test case. DiffOPF captures the modality and support of the underlying distribution of dispatch setpoints, while single-valued DeepOPF only provides point predictions.}
    \label{fig:DiffOPF_vs_DeepOPF}
\end{figure}

\begin{figure}
    \centering
    \includegraphics[width=1\linewidth]{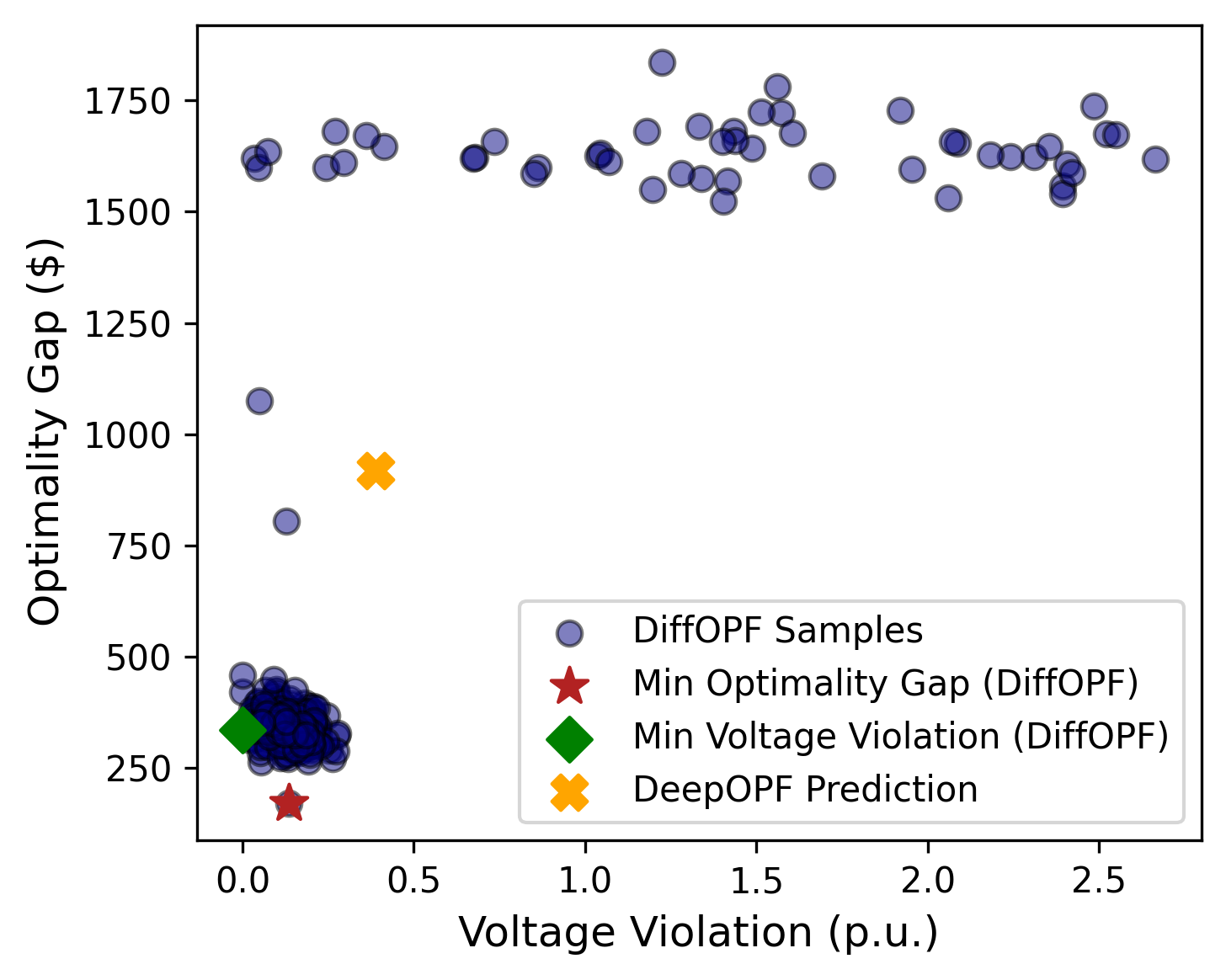}
    \caption{Trade-off between optimality gap and voltage constraint violations for DiffOPF warm starts on a single load instance in the PJM test case.}
    \label{fig:VoltageVsOptGap}
\end{figure}

\begin{table*}[ht]
\setlength{\tabcolsep}{12pt}
\centering
\caption{Comparison of warm start performance between DiffOPF and DeepOPF across the three metrics defined in \eqref{metrics}.}
\label{tab:warmstart_quality}
\begin{tabular}{llrrcccc}
\toprule
\multirow{2}{*}{Test Case} & \multirow{2}{*}{Solver} & \multicolumn{2}{c}{Optimality gap [$\$$]} & \multicolumn{2}{c}{Setpoint error [$\text{p.u.}^{2}$]} & \multicolumn{2}{c}{Voltage violation [p.u.]} \\
\cmidrule(lr){3-4} \cmidrule(lr){5-6} \cmidrule(lr){7-8}
 &  & \multicolumn{1}{c}{Mean} & \multicolumn{1}{c}{$\text{CVaR}_{10\%}$}  & \multicolumn{1}{c}{Mean} & \multicolumn{1}{c}{$\text{CVaR}_{10\%}$} & \multicolumn{1}{c}{Mean} & \multicolumn{1}{c}{$\text{CVaR}_{10\%}$} \\
\midrule
\multirow{2}{*}{PJM} 
 & DiffOPF  & 8.0 (0.1$\%$) &46.1 (0.3$\%$) &0.000  &0.000 &0.132  & 0.310 \\
 & DeepOPF  & 1091.3 (8.6$\%$) & 4134.4 (33.5$\%$) & 0.004 & 0.012 & 0.389 & 0.680 \\
\midrule
\multirow{2}{*}{PJM-API} 
 & DiffOPF  & 504.0 (0.7$\%$) & 1246.5 (1.9$\%$) & 0.009 & 0.019 & 0.002 & 0.022 \\
 & DeepOPF  & 1142.1 (1.7$\%$) & 2873.2 (4.3$\%$) & 0.044 & 0.160 & 0.264 & 1.191 \\
\midrule
\multirow{2}{*}{IEEE 14-Bus} 
 & DiffOPF  & 3.4 (0.2$\%$) & 13.3 (0.6$\%$) & 0.000 & 0.000 & 0.002 & 0.011 \\
 & DeepOPF  & 4.6 (0.2$\%$)  & 15.5 (0.7$\%$) & 0.000& 0.000 & 0.004 &0.013  \\
\midrule
\multirow{2}{*}{IEEE 24-Bus} 
 & DiffOPF  & 1.8 (0.0$\%$) & 5.7 (0.0$\%$) & 0.000 & 0.000 & 0.004 & 0.016 \\
 & DeepOPF  & 290.3 (0.6$\%$) & 718.8 (1.4$\%$) & 0.000 & 0.002 & 0.017 & 0.038 \\
\midrule
\multirow{2}{*}{IEEE 118-Bus} 
 & DiffOPF  & 95.3 (0.1$\%$) & 378.1 (0.5 $\%$) & 0.000 & 0.000 & 0.007 & 0.036 \\
 & DeepOPF  & 8620.5 (10.9$\%$) & 18333.4 (23.8 $\%$) & 0.350 & 3.500 & 1.087 & 4.430 \\
\bottomrule
\end{tabular}
\end{table*}

\subsection{Tests Across Benchmark Power Systems}\label{subsec:tests_bench}

Table~\ref{tab:warmstart_quality} summarizes the outcomes of the projection problem \eqref{eq:pp_opt} that receives warm starts from DeepOPF and DiffOPF models for selected benchmark power systems. The table provides statistics across 100 load instances for the three metrics defined in \eqref{metrics}, including the mean values over 100 load instances and the mean of 10 worst-case instances, represented by the conditional value at risk at the 10th quantile $\text{CVaR}_{10\%}$.

Across all test cases, DiffOPF consistently outperforms DeepOPF in terms of optimality gap, setpoint error, and voltage violation, both on average and under worst-case conditions. For instance, in the IEEE 118-bus system, DiffOPF achieves an average optimality gap of $0.1\%$ compared to $10.9\%$ for DeepOPF, and reduces the mean voltage violation from $1.087~\text{p.u.}$ to $0.007~\text{p.u}$. As another example, in the IEEE 14-bus system, the performance of DiffOPF and DeepOPF is very similar. This is because only two generators contribute to active power, and one of them, being significantly cheaper and having larger capacity, almost always supplies all of the load, thus limiting the multi-valued nature of the OPF mapping.

\subsection{Sample Complexity}
We validate the sample complexity requirement of DiffOPF introduced in Sec.~\ref{subsec:sample_comp}, and report the number of samples required to obtain at least one $\varepsilon$-close solution at different confidence levels $\delta$ for the smallest PJM $5$-bus and the largest IEEE $118$-bus test cases. To compute $p_{\varepsilon}$, we consider the same $100$ test load instances as in Sec.~\ref{subsec:tests_bench}. For each load instance, we generate $250$ samples using DiffOPF and count the number of samples that are $\varepsilon$-close to the solution obtained by the conventional interior-point method. The empirical probability $p_{\varepsilon}$ is computed as the fraction of $\varepsilon$-close samples among the total $100 \times 250$ generated samples.

Table~\ref{tab:sample_complexity} reports the theoretical lower bound $M$ on the number of samples required to obtain at least one $\varepsilon$-close solution with confidence level $\delta$, together with the empirical number $N_{\varepsilon}$ of $\varepsilon$-close samples observed among the $M$ generated samples. Comparing $N_{\varepsilon}$ with $M$ provides insight into the conservativeness of the theoretical bound. Note that we present the distance $\varepsilon$ in percentage, as the cost structures of the two systems are different.

The lower bound appears conservative for the small PJM $5$-bus system, where a relatively large fraction of the $M$ samples are $\varepsilon$-close to the optimal solution (i.e., $N_{\varepsilon}$ is often comparable to $M$). In contrast, for the larger IEEE $118$-bus system the bound becomes noticeably tighter, with $N_{\varepsilon}$ remaining small relative to $M$. As $\varepsilon$ increases, fewer samples are required to obtain an $\varepsilon$-close solution. Even for a small tolerance such as $\varepsilon = 0.5\%$, DiffOPF requires only a modest number of samples, not exceeding $269$.
\begin{table}[ht]
\centering
\caption{Sample complexity of DiffOPF across benchmark test cases: theoretical lower bound $M$ and the actual number $N_\varepsilon$ of $\varepsilon$-close samples for different confidence levels $\delta$.}
\label{tab:sample_complexity}
\begin{tabular}{l l c c c c c c c}
\toprule
Test Case & $\varepsilon~\%$ & $p_{\varepsilon}$ & \multicolumn{2}{c}{$\delta=0.9$} & \multicolumn{2}{c}{$\delta=0.95$} & \multicolumn{2}{c}{$\delta=0.99$} \\
\cmidrule(lr){4-5} \cmidrule(lr){6-7} \cmidrule(lr){8-9}
 &  &  & $M$ & $N_\varepsilon$ & $M$ & $N_\varepsilon$ & $M$ & $N_\varepsilon$ \\
\midrule
\multirow{4}{*}{PJM 5-Bus}
& $0.5$ & 0.07 & 32 & 28 & 41 & 27 & 63 & 45 \\
& $1$   & 0.49 & 4 & 4 & 5 & 4 & 7 & 5 \\
& $2$   & 0.77 & 2 & 2 & 3 & 3 & 4 & 4 \\
& $5$   & 0.94 & 1 & 1 & 1 & 1 & 1 & 1 \\
\midrule
\multirow{4}{*}{IEEE 118-Bus}
& $0.5$ & 0.02  & 135 & 1 & 175 & 3 & 269 & 2 \\
& $1$   & 0.03 & 71 & 3  & 93 &3  & 142 & 4 \\
& $2$   & 0.05 & 42 &2 & 54 & 3 & 83 & 4 \\
& $5$   & 0.17 & 13 & 1 & 17 &5 & 26 & 2 \\
\bottomrule
\end{tabular}
\end{table}

\section{Conclusion}

We developed a diffusion-based solver, DiffOPF, that successfully approximates the multi-valued load-to-dispatch mapping in AC-OPF, and analyzed its sample complexity to inform applications across small and large power systems. Compared to single-valued solvers, DiffOPF allows system operators to choose from a diverse set of warm starts, favoring smaller voltage violations, lower dispatch costs, or a combination of these and other operational objectives. 

For future work, we plan to extend the proposed framework beyond warm start applications by explicitly incorporating OPF constraints within the diffusion sampling process. Building on our prior work in~\cite{hoseinpour2025constraineddiffusionmodelssynthesizing}, where power flow constraints were enforced through gradient guidance, we aim to integrate OPF constraints to ensure the feasibility of generated solutions. In addition, we intend to address the runtime gap between DiffOPF, which takes at most $2.3$ seconds for the IEEE $118$-bus system, and single-valued deep learning OPF solvers that achieve millisecond runtimes. To this end, we will explore accelerating DiffOPF using diffusion distillation techniques, which compress the sampling process into only a few denoising steps~\cite{meng2023distillation}.

\balance
\bibliographystyle{IEEEtran}
\bibliography{mybib.bib}

\endgroup
\balance
\end{document}